\documentclass[11pt]{article}
\usepackage{amsmath, amssymb, epsfig,psfrag}
\usepackage{epsf,citesort}
\newcommand{\cN}{\mathcal{N}}

\newcommand{\cI}{\mathcal{I}}

\newcommand{\beps}{\varepsilon} 
\newcommand{\bx}{{\mathbf x}}
\newcommand{\by}{{\mathbf y}}
\newcommand{\IP}{{\sf IP}}
\newcommand{\LP}{{\sf LP}}
\newcommand{\DL}{{\sf DUAL}}

\newcommand{\lf}{\left}
\newcommand{\rf}{\right}
\newcommand{\bone}{\mathbf{1}}
\newcommand{\bQ}{\mathbb Q}

\newcommand{\beq}{\begin{eqnarray}}
\newcommand{\eeq}{\end{eqnarray}}
\newcommand{\beqn}{\begin{eqnarray*}}
\newcommand{\eeqn}{\end{eqnarray*}}

\newcommand{\bL}{{\mathbf L}}
\newcommand{\bW}{{\mathbf W}}
\newcommand{\bzero}{{\mathbf 0}}
\newcommand{\cT}{{\cal T}}

\newcommand{\bT}{{\mathbb T}}

\newcommand{\hI}{\hat{I}}

\renewcommand{\epsilon}{\varepsilon}

\newcommand{\eproof}{\hfill $\blacksquare$}
\newtheorem{theorem}{Theorem}[section]
\newtheorem{lemma}{Lemma}[section]

\newtheorem{corollary}{Corollary}[section]

\title{Tightness of LP via Max-product Belief Propagation}

\author{
Sujay Sanghavi \quad \quad \quad \quad Devavrat Shah\footnote{Laboratory for Information
and Decision Systems, MIT. Email: {\tt \{sanghavi, devavrat\}@mit.edu} }
}

\date{}

\begin{document}

\maketitle

\begin{abstract}
We investigate the question of tightness of linear programming (LP)
relaxation for finding a maximum weight independent set (MWIS) in 
sparse random weighted graphs. We show that an edge-based LP relaxation is 
asymptotically tight for Erdos-Renyi graph $G(n,c/n)$ for $c \leq 2e$ 
and random regular graph $G(n,r)$ for $r\leq 4$ when node weights
are i.i.d. with exponential distribution of mean $1$. We establish 
these results, through a precise relation between the tightness of 
LP relaxation and convergence of the max-product belief propagation 
algorithm. We believe that this novel method of understanding 
structural properties of combinatorial problems through properties
of iterative procedure such as the max-product should be of interest 
in its own right. 
\end{abstract}

\section{Introduction}

The max-weight independent set (MWIS) problem is the following: given
a graph with weights on the nodes, find the heaviest set of disjoint
nodes. It is a canonical combinatorial optimization problem, known to
be NP-hard \cite{Kar72} and hard to approximate \cite{Has92} in the
worst case. This has led to considerable interest in average-case
characterizations of fundamental structural properties, as well as the
hardness, of the MWIS problem. We summarize some of this work in Section
\ref{ssec:rel-work}.

In this paper we establish that, with high probability, the standard
simple edge-based linear programming (LP) relaxation of the MWIS
problem is asymptotically tight on $1-o(1)$ fraction of the nodes, for
the Erdos-Renyi graph $G(n,c/n)$ for $c \leq 2e$ and random regular
graph $G(n,r)$ for $r\leq 4$, when the node weights are drawn
i.i.d. with exponential distribution of mean $1$. This means that
problems from this ensemble are ``easy'' with high probability, since
the LP can be solved in polynomial time \cite{GLS84}.

We arrive at this result via an analysis of the max-product form of
belief propagation. In particular, we establish the following two
properties of max-product: {\em (a)} for any {\em arbitrary} problem
instance, max-product succeeds\footnote{That is, the estimate of max-product 
converges, which may not necessarily correct.} for a given node 
{\em only if} every LP optimum assignment is integral for that node, 
and {\em (b)} for the random weighted graphs above, max-product succeeds for almost all nodes, with
high probability. To the best of our knowledge, our work represents
the first instance where analysis of iterative procedures like
max-product has been used as a tool to establish fundamental
properties of optimization problems on graphs; usually, the analysis
goes the other way. We believe that this method of analysis has the
potential to shed insight into other graph-theoretic/algorithmic
problems beyond the ones presented in this paper.

Other motivations for our work are: {\em (1)} to obtain a better
understanding of iterative procedures like max-product and its relation
to underlying problem structure, a topic of
much recent interest (for example, \cite{x1,x2,x3,x4,x5}, and {\em (2)} to
characterize the performance of max-product as a simple, distributed
solution to the MWIS problem, in applications where such a solution is
needed (e.g. wireless scheduling \cite{JS07}).

We now summarize some of the most closely related work on average-case
analysis of graph problems (in particular MWIS), and on the analysis
of max-product. Then we summarize our main contributions, and provide
an outline for the rest of the paper.

\subsection{Related Work}\label{ssec:rel-work}

We now summarize the most relevant existing literature in the two
areas concerning this paper: average-case analysis of independent set
problems, and analysis of max-product and belief propagation.

There has been much work on average-case characterizations of hard
problems, see \cite{Frieze-Mcdiarmid,Trevisan_average}. For the
(unweighted) independent set problem, there has been much work on the
dense Erdos-Renyi graphs $G(n,\frac{1}{2})$, where it is known that
the max-size independent set is of size $2\log_2 n$ almost
surely. However, no algorithm is known to find efficiently an
independent set of size significantly larger than $\log_2 n$,
\cite{Frieze-Mcdiarmid,karp_probab}.  Feige \cite{feige}, and Feige
and Krauthgamer \cite{feige-kraut} investigate the tightness of the
Lovasz-Schriver hierarchy of relaxations for these graphs -- relaxation
upto $\log n$ level will lead to $1-o(1)$ approximation. In constrast,
here we prove tightness of LP relaxation for sparse random graphs.

There has also been substantial interest in independent set 
problems on sparse random graphs. Karp and Sipser \cite{karp_sipser} analyze a greedy
algorithm for matchings on sparse Erdos-Renyi graphs $G(n,c/n)$; a
similar analysis for independent set shows that the algorithm works
for $c\leq e$ (see e.g. \cite{GAM05} for details).  Frieze and Suen
\cite{frieze-suen} investigate the success of a greedy algorithm for
independent sets in random 3-regular graphs. Bollabas \cite{bollabas}
provides an upper bound on the size of the largest independent set for
regular graphs.

There is a tremendous amount of literature on the study of
message-passing algorithms for inference. We will now briefly review
the existing work most directly related to this paper. The main
results of this paper illustrate a close relationship between the
performance of the max-product algorithm, and linear programming. Such
precise or semantic connections have been noticed in other contexts:
in decoding of linear codes \cite{vontobel}, weighted matching
\cite{BSS07,sanghavi,bayati-msr} and weighted independent set
\cite{SSW07}. For more general inference problems, several authors
\cite{WJ03,kolmogorov,weiss07,globerson} develop alternative
message-passing algorithms that explicitly solve linear programming
relaxations. These are related to, but not the same as, the classical
max-product belief propagation that we will use in this paper.

\subsection{Our Contributions and Paper Organization}

In this paper we investigate the (simple, edge-based) LP relaxation of
the MWIS problem for certain random ensembles. The weights on the
nodes in our random ensembles are drawn from a continuous
distribution, and hence the LP optimum $x^*$ will be unique with
probability 1. In general, each node $i$ will be assigned a (possibly
fractional\footnote{In fact, it is known \cite[Theorem 64.7]{schrijver}
that the edge-based LP is half-integral: $x_i^* = 0,1$ or
$\frac{1}{2}$}) value $x^*_i$ at this optimum. Our main result in this
paper is the following theorem.

\begin{theorem}\label{thm:main}
Let $G$ be an Erdos-Renyi random graph $G(n,c/n)$ for $c \leq 2e$, or
random regular graph $G(n,r)$ for $r\leq 4$. Suppose each node has a
weight that is chosen to be i.i.d. with exponential distribution of
mean $1$, independent of the graph. Let $x^*$ be the optimum of the
edge-based LP relaxation of the MWIS problem (described in section
\ref{sec:prelim}). Let $i$ be a node picked uniformly at random,
independent of everything else\footnote{Or, alternatively, one can
think of having an a-priori node numbering before the edges and
weights are picked.}. Then, given any $\beps$, there exists an
$N(\beps)$ such that $\Pr[x_i^* = \text{0 or 1}] \geq 1-\beps$ as
long as $n\geq N(\beps)$.
\end{theorem}

The above theorem states that LP will be tight on a fraction $1-o(1)$
of the nodes, with high probability. We arrive at this result via the
novel route of max-product belief propagation. Max-product is an
iterative algorithm, and at every iteration produces an estimate
$\hat{x}_i^t = 0,1$ or ? (corresponding to ``not in the MWIS'', ``in
the MWIS'', and ``don't know'' respectively) for each node $i$ and time
$t$. In this paper, we prove the following two results on the
performance of max-product:
\begin{enumerate}
\item For any {\em arbitrary} graph with arbitrary weights, consider a
  node $i$. If there exists {\em any} LP optimum $x^*$ that puts a
  fractional mass $x^*_i$ on node $i$, then the max-product estimate
  will be $\hat{x}^t_i = 0$ or ? for every odd time $t$, and
  $\hat{x}^t_i = 1$ or ? for every even time $t$. This is Theorem
  \ref{thm:lp_mp}.
\item For the random graph ensembles, and randomly chosen node $i$ as
  in Theorem \ref{thm:main} above, given any $\beps>0$ there exists an
  $N(\beps)$ and $t(\beps)$ such that for all $n\geq N(\beps)$, the
  max-product estimate $\hat{x}^t_i$ remains constant and equal to
  either 0 or 1, for all $t\geq t(\beps)$. This is Theorem
  \ref{thm:main1}.
\end{enumerate}

Note that the first result (Theorem \ref{thm:lp_mp}) is
non-asymptotic: it holds for finite graphs and number of iterations
(i.e. it is not a ``fixed point'' analysis). This is crucial in
avoiding an ``order of limits'' problem (between the number of
iterations and the size of the problem) that may otherwise come up in
establishing the overall result. The second result is established
using ideas from the method of local weak convergence
\cite{aldous,GAM05}.

In addition to the fact that they together immediately imply our main
result above, we believe that each of the two theorems above are
interesting in their own right. The first theorem generalizes to the
case when ``clique factors'' are added to max-product, and clique
constraints to the corresponding LP. However, in this paper we will
concentrate only on the simplest edge-based case.

The theorems also shed light on the usefulness of max-product as a
distributed heuristic for the MWIS problem. Consider first an
arbitrary graph. In light of Theorem \ref{thm:lp_mp}, we can stop
max-product after a certain number of iterations and check for
one-step agreement: if the estimate $\hat{x}^t_i = \hat{x}^{t+1}_i =
0$ or 1, then we know that $x^*_i = 1$ for any LP optimum. This is
then also consistent with the MWIS (see Lemma \ref{lem:partial}
below). This means that the set of nodes for which $\hat{x}^t_i =
\hat{x}^{t+1}_i = 1$ will form an independent set; we can stop
max-product at any time and have a candidate independent set (although
it may not be the MWIS). If we restrict our attention to the ensembles
considered above, if $n \geq N(\beps)$ and max-product is run for a
sufficient number of iterations $t\geq t(\beps)$, this set obtained
from one-step agreement will be pretty large in size. The fact that
the weights come from an exponential distribution means that this
candidate independent set will be a very good approximation of the
MWIS. This is shown in Theorem \ref{cor:main1}.

The rest of the paper is organized as follows. In Section
\ref{sec:prelim} we lay out the groundwork and preliminaries. We also
describe precisely the max-product algorithm we are considering in
this paper, and state some of its other known properties. In Section
\ref{sec:one} we state and prove Theorem \ref{thm:lp_mp}. In Section
\ref{sec:two} we state and prove Theorem \ref{thm:main1}, and
Theorem \ref{cor:main1}.

\section{Preliminaries: MWIS, LP relaxation and max-product}\label{sec:prelim}

Consider a graph $G = (V,E)$, with a set $V$ of nodes and a set $E$ of
edges. Let $\cN(i)=\{j \in V: (i,j) \in E\}$ be the neighbors of $i
\in V$.  Positive weights $w_i, i\in V$ are associated with each node.
A subset of $V$ will be represented by vector $\bx = (x_i) \in
\{0,1\}^{|V|}$, where $x_i = 1$ means $i$ is in the subset $x_i = 0$
means $i$ is not in the subset. A subset $\bx$ is called an {\em
independent set} if no two nodes in the subset are connected by an
edge: $(x_i, x_j) \neq (1,1)$ for all $(i,j) \in E$. We are interested
in finding a maximum weight independent set (MWIS) $\bx^*$.  This can
be naturally posed as an integer program, denoted below by \IP.~ The
{\em linear programing relaxation}\footnote{Other (tighter) LP
relaxations of \IP~ are possible, and some of our results carry over
to those relaxations as well. However, in this paper we will
concentrate only on the LP relaxation presented above.} of \IP~ is
obtained by replacing the integrality constraints $x_i\in \{0,1\}$
with the constraints $x_i\geq 0$. We will denote the corresponding
linear program by \LP.~The dual of \LP~ is denoted below by \DL.
{\small \begin{align*}
& \IP : ~{\sf max} ~~ \sum_{i=1}^n w_i x_i \qquad \mbox{over}~x_i\in \{0,1\}, & ~& ~\qquad\qquad \LP : ~{\sf max} ~~ \sum_{i=1}^n w_i x_i, \qquad \mbox{over}~x_i \geq 0, \\
& {\sf s.t.} ~~ x_i + x_j \leq 1 ~~\text{for all}~ (i,j) \in E, &~&~ \qquad\qquad
{\sf s.t.} ~~x_i + x_j \leq 1 ~~\text{for all}~ (i,j) \in E, 
\end{align*}}
It is well-known that \LP~ can be solved efficiently, and if it has an
integral optimal solution then this solution is an MWIS of $G$. If
this is the case, we say that there is no {\em integrality gap}
between \LP~ and \IP~ or equivalently that the \LP~relaxation is {\em
tight}.  We refer an interested reader to book by Schrijver 
\cite{schrijver} for many interesting properties of the 
\LP. We note one property: partial correctness.

\begin{lemma}[\cite{schrijver},Corollary 64.9a)]\label{lem:partial} LP
optima are partially correct: for any graph, any LP optimum $x^*$
and any node i, if the mass $x^*_i$ is integral then there exists an
MWIS of $G$ for which $i$'s membership is given by $x^*_i$.
\end{lemma}

The classical max-product algorithm is a heuristic that can be used to
find the MAP assignment of a probability distribution. Before, we state
the (simplified) max-product algorithm applied to the MWIS problem, we
state probability distribution whose MAP solution corresponds to solution
of MWIS problem for completeness. Now, given an MWIS problem on $G=(V,E)$, 
associate a binary random variable $X_i$ with each $i \in V$ and 
consider the following joint distribution: for $\bx \in \{0,1\}^n$,
\begin{eqnarray}
p\left(\bx\right ) & = & \frac{1}{Z} \prod_{(i,j) \in E}
\mathbf{1}_{\{x_i + x_j \leq 1\}} \prod_{i \in V} \exp(w_i x_i),
\label{e1}
\end{eqnarray}
where $Z$ is the normalization constant. In the above, $\mathbf{1}$ is
the standard indicator function: $\bone_{\text{true}} = 1$ and
$\bone_{\text{false}}=0$. It is easy to see that $p(\bx) =
\frac{1}{Z}\exp\lf(\sum_i w_i x_i\rf)$ if $\bx$ is an independent set,
and $p(\bx) = 0$ otherwise.  Thus, any MAP estimate $\arg\max_{\bx}
p(\bx)$ corresponds to a maximum weight independent set of $G$.

Here, we present a simplified  version of the max-product algorithm
(obtained by taking logarithm of ratio of messages for the original
algorithm) -- we refer an interested reader to \cite{SSW07} for details
on the transformation. The algorithm is iterative; in iteration $t$ 
each node $i$ sends a {\em message} $\{\gamma^{t}_{i\to j}\}$ to each 
neighbor $j\in \cN(i)$ based on $\{\gamma^t_{k\to i}\}, k \in \cN(i)$.  
Each node $i$ also maintains an {\em estimate} of its assignment in 
independent set $\{x_i(\gamma^t)\}$ based on messages it received $\{\gamma^t_{j\to i}\}$,
$j \in \cN(i)$. The following describes the message and estimate updates, 
as well as the final output.

\vspace{.05in}
\noindent{\bf Max-product for MWIS}
\vspace{.05in}
\hrule
\begin{itemize}
\item[(o)] Initially, $t = 1$ and $\gamma^1_{i\to j} = 0$ for
  all $(i,j) \in E$.
\vspace{-.1in}
\item[(i)] The messages are updated as follows: for $t > 1,$
{\small \begin{eqnarray}
\gamma^{t}_{i\rightarrow j} = \left ( w_i - \sum_{k \in
\cN(i)-j}\gamma^{t-1}_{k\rightarrow i} \right )_+, \label{eq:min-sum}
\end{eqnarray} }
\vspace{-.1in}
\item[(ii)] Estimate max. wt. independent set $\bx(\gamma^{t})$ as
follows:
\begin{eqnarray}
x_i(\gamma^t) & = & \begin{cases} 1 
   & \qquad \text{if} \qquad w_i > \sum_{k\in \cN(i)}  \gamma^t_{k\rightarrow i} \label{eq:est_1_def} \\ 
                                  0  
   & \qquad \text{if} \qquad w_i < \sum_{k\in \cN(i)} \gamma^t_{k\rightarrow i} \\
                                  ? 
   & \qquad \text{if} \qquad  w_i = \sum_{k\in \cN(i)} \gamma^t_{k\rightarrow  i}.
   \end{cases}
   \end{eqnarray} 

\item[(iv)] Update $t = t + 1$; repeat from (i)

\end{itemize}
\vspace{.1in}
\hrule

\subsection{Max-product: known properties}

A popular technique for analyzing max-product is to consider its fixed
points \cite{weiss_freeman,aji_mceliece,yedidia}. Here, we list
relevant properties of the max-product for MWIS that are established
in \cite{SSW07} for setting up context for the results stated in
Section \ref{sec:one}. Note that a set of messages $\gamma^*$ is a
fixed point of max-product if, for all $(i,j)\in E$,
\begin{equation}
\gamma^*_{i\rightarrow j} = \left ( w_i - \sum_{k \in \cN(i)-
  j}\gamma^*_{k\rightarrow i} \right )_+ \label{eq:fixedp}
\end{equation}
The following is a summary of results from \cite{SSW07}. 
\begin{theorem}{\cite{SSW07}}\label{thm:fp}
There exists at least one fixed point $\gamma^*$ such that
$\gamma^*_{i\to j} \in [0,w_i]$ for each $(i,j)\in E$. Given
such a fixed point $\gamma^*$, let $\bx(\gamma^*) =
(x_i(\gamma^*))$ be the corresponding estimate. Define
$\by = (y_i) \in [0,1]^n$ as follows: $y_i = \frac{1}{2}$\; if
$x_i(\gamma^*) = ?$,\; and $y_i = x_i(\gamma^*)$ otherwise.
Then, $\by$ corresponds to an extreme point of the LP for
MWIS.
\end{theorem}
Theorem \ref{thm:fp} implies that the fixed point estimate
of max-product for MWIS is an extreme point of \LP, and hence
one that maximizes {\em some} weight function consisting of 
positive node weights. Note however that this {\em may not} be the 
true weights $w_i$. In other words, given any MWIS problem 
with graph $G$ and weights $w$, each max-product fixed
point represents the optimum of the LP relaxation of some MWIS problem
on the same graph $G$, but possibly with different weights
$\widehat{w}$. The fact that max-product estimates optimize a 
different weight function means that both eventualities are 
possible: \LP~ giving the correct answer but max-product failing, 
and vice versa. In \cite{SSW07}, two examples are presented
for each one of these situations. These examples indicate 
that it may not be possible to resolve the question of 
relative strength of the two procedures based solely on an
analysis of the fixed points of max-product.

\section{Max-product: Convergence and tightness of \LP}
\label{sec:one}

The max-product is a deterministic iterative algorithm.
It may have multiple fixed points. In which case, it 
converges (if it does) to a fixed point depending upon
its starting condition. In absence of prior information, 
the ``natural'' (and popular) initialization of messages
is the one described in Section \ref{sec:prelim} (i.e. 
$\gamma_{i\to j}^0 = 0,$ for all $(i,j) \in E$. In this
section, we directly analyze the performance of max-product
algorithm with these initial conditions.  We show that the 
resulting estimates are very exactly characterized by optima 
of the {\em true} \LP, at every time instant (not just at fixed
points). This implies that, if a fixed point is reached, it will
exactly reflect an optimum of \LP. Our main theorem for this section is
stated below.

\begin{theorem}\label{thm:lp_mp}
Given any MWIS problem on weighted graph $G$, suppose max-product is
started from the initial condition $\gamma = 0$. Then, for any node
$i\in G$.
\begin{enumerate}
\item If there exists {\em any} optimum $x^*$ of ~\LP~ for which the
  mass assigned to edge $i$ satisfies $x^*_i >0$, then the max-product
  estimate $x_i(\gamma^t)$ is 1 or ? for all {\em odd} times $t$.

\item If there exists {\em any} optimum $x^*$ of ~\LP~ for which the
  mass assigned to $i$ satisfies $x^*_i <1$, then the max-product
  estimate $x_i(\gamma^t)$ is 0 or ? for all {\em even} times $t$.
\end{enumerate}
\end{theorem}

An important and direct consequence of the above result is as follows.
\begin{corollary}\label{cor:lp_mp}
If \LP~ has non-integral optima, i.e. there is $i$ such that $x^*_i \in (0,1)$.
Then, estimate $x_i(\gamma^t)$ either oscillates or converges to ?. 
\end{corollary}

The proof of this theorem relies on the computation tree
interpretation of max-product estimates. We now specify this
interpretation for our problem, and then prove Theorem
\ref{thm:lp_mp}.

\subsection{Computation Tree for MWIS}

\begin{figure}[h]
\centering{\epsfig{file=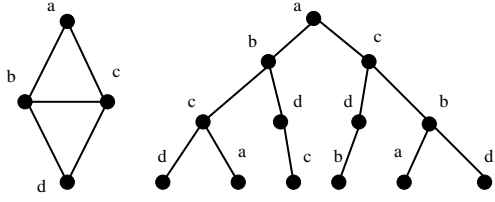,height = 1.0in}}
\caption{An example of computation tree for $t=4$ iterations for a $4$ node graph.}
\label{fig:mwis_comp_tree}
\end{figure}

The proof of Theorem \ref{thm:lp_mp} relies on the well-known 
computation tree interpretation 
\cite{weiss_freeman,bib:tatik_jordan} of the
loopy max-product estimates. In this section we briefly outline this
interpretation.  For any node $i$, the {\em computation tree} at time
$t$, denoted by $T_i(t)$, is defined recursively as follows: $T_i(1)$
is just the node $i$. This is the {\em root} of the tree, and in this
case is also its only leaf. The tree $T_i(t)$ at time $t$ is generated
from $T_i(t-1)$ by adding to each leaf of $T_i(t-1)$ a copy of each of
its neighbors in $G$, {\em except for the one neighbor that is already
present in $T_i(t-1)$}. Each node in $T_i$ is a copy of a node in $G$,
and the weights of the nodes in $T_i$ are the same as the
corresponding nodes in $G$. As an example, Figure \ref{fig:mwis_comp_tree}
presents computation tree $T_a(4)$ for the node $a$ in graph $G$ at time
$t=4$. 

\begin{lemma}\label{lem:comp_tree}
For any node $i$ at time $t$,
(a) $x_i(\gamma^t) = 1$ if and only if the root of $T_i(t)$ is a
  member of {\em every} MWIS on $T_i(t)$; 
  (b) $x_i(\gamma^t) = 0$ if and only if the root of $T_i(t)$ is not a
  member of {\em any} MWIS on $T_i(t)$; and (c) $x_i(\gamma^t) = ?$ else.
\end{lemma}
Thus the max-product estimates correspond to max-weight independent
sets on the computation trees $T_i(t)$, as opposed to on the original
graph $G$. 
 
\subsection{Proof of Theorem \ref{thm:lp_mp}}

We now prove Theorem \ref{thm:lp_mp}. We will present the 
proof of part (1). The proof of part (2) follows from very similar
arguments and hence we will skip it. 

We will prove part 1 through contradiction. To this end, suppose part
1 is not true. That is, there exists node $i$, an optimum $x^*$ of
\LP~ with $x^*_i>0$, and an odd time $t$ at which the estimate is
$\hat{x}^t_i = 0$. For brevity, in the remainder of the proof we will
use the notation $\hat{x}^t_i = x_i(\gamma^t)$ for the estimates.  Let
$T_i(t)$ be the corresponding computation tree. Using Lemma
\ref{lem:comp_tree} this means that the root $i$ is {\em not} a member
of any MWIS of $T_i(t)$. Let $I$ be some MWIS on $T_i(t)$. We now
define the following set of nodes
\begin{equation*}
I^* = \left \{ j\in T_i(t) \,:\,j\notin I, ~\text{and copy of $j$ in
  $G$ has $x^*_{j}>0$}  \right \} \label{eq:istar}
\end{equation*}
In words, $I^*$ is the set of nodes in $T_i(t)$ which are not in
$I$, and whose copies in $G$ are assigned strictly positive mass by
the LP optimum $x^*$.

Note that by assumption the root $i\in I^*$ and $i\notin I$. Now, from
the root, recursively build a {\em maximal alternating subtree} $S$ as
follows: first add root $i$, which is in $I^*-I$. Then add all
neighbors of $i$ that are in $I-I^*$. Then add all {\em their}
neighbors in $I^*-I$, and so on. The building of $S$ stops either when
it hits the bottom level of the tree, or when no more nodes can be
added while still maintaining the alternating structure.  Note the
following properties of $S$:
\begin{itemize}
\item $S$ is the disjoint union of $(S\cap I)$ and $(S\cap I^*)$.
\item For every $j\in S\cap I$, all its neighbors in $I^*$ are
  included in $S\cap I^*$. Similarly for every $j\in S\cap I^*$, all
  its neighbors in $I$ are included in $S\cap I$.
\item Any edge $(j,k)$ in $T_i(t)$ has at most one endpoint in $(S\cap
  I)$, and at most one in $(S\cap I^*)$.
\end{itemize}
We now state a lemma, which we will prove later. The proof uses the
fact that $t$ is odd.

\begin{lemma}\label{lem:set_flip_lp_mp}
  The weights satisfy $w(S\cap I) \leq w(S\cap I^*)$.
\end{lemma}

We now use this lemma to prove the theorem. Consider the set $I'$
which changes $I$ by flipping $S$:
\[
I' = I - (S\cap I) + (S\cap I^*)
\]
We first show that $I'$ is also an independent set on $T_i(t)$. This
means that we need to show that every edge $(j,k)$ in $T_i(t)$ touches
at most one node in $I'$. There are thus three possible scenarios for
edge $(j,k)$:
\begin{itemize}
\item $j,k \notin S$. In this case, membership of $j,k$ in $I'$ is the
  same as in $I$, which is an independent set. So $(j,k)$ has at most
  one node touching $I'$.
\item One node $j\in S\cap I$. In this case, $j\notin I'$, and hence
  again at most one of $j,k$ belongs to $I'$.
\item One node $k\in S\cap I^*$ but other node $j\notin S\cap I$. This
  means that $j\notin I$, because every neighbor of $k$ in $I$ should
  be included in $S\cap I$. This means that $j\notin I'$, and hence
  only node $k\in I'$ for edge $(j,k)$.
\end{itemize}
Thus $I'$ is an independent set on $T_i(t)$. Also, by Lemma
\ref{lem:set_flip_lp_mp}, we have that
\[
w(I')~ \geq ~ w(I)
\]
However, $I$ is an MWIS, and hence it follows that $I'$ is also an
MWIS of $T_i(t)$. However, by construction, root $i\in I'$, which
violates the fact that $\hat{x}_i(t) = 0$. The contradiction is thus
established, and part (1) of the theorem is proved. 
\hfill $\blacksquare$

\subsubsection{Remaining proofs}

The proof of this lemma involves a perturbation argument on the
\LP. For each node $j\in G$, let $m_{j}$ denote the number of times
$j$ appears in $S\cap I$ and $n_{j}$ the number of times it appears in
$S\cap I^*$. For $\epsilon > 0$, define
\begin{equation}
x ~ = ~ x^* + \epsilon (m - n). \label{eq:x}
\end{equation}
We now show state a lemma that is proved immediately following this
one.
\begin{lemma} \label{lem:feas}
$x$ is a feasible point for \LP, for small enough $\epsilon$.
\end{lemma}
Using the above, we will complete the proof of
Lemma \ref{lem:set_flip_lp_mp}. Since $x^*$ is an optimum of \LP, it follows
that $w'x \leq w'x^*$, and so $w'm \leq w'n$. However, by definition,
$w'm = w(S\cap I)$ and $w'n = w(S\cap I^*)$. This finishes the
proof of Lemma \ref{lem:set_flip_lp_mp}. \hfill $\blacksquare$

\noindent {\em Proof of Lemma \ref{lem:feas}:}
Now, we complete proof of the only remaining part: Lemma \ref{lem:feas}. 
We wish to show that  $x$ as defined in (\ref{eq:x}) is a feasible
point for \LP, for small enough $\epsilon > 0$. To do so we have to check
node constraints $ x_{j}\geq 0$ and edge constraints $x_j + x_k \leq
1$ for every edge $(j,k)\in G$. Consider first the node
constraints. Clearly we only need to check them for any $j$ which has
a copy $j\in I^*\cap S$. If this is so, then by the definition
(\ref{eq:istar}) of $I^*$, $x_{j}^* >0$. Thus, for any $m_{j}$ and
$n_{j}$, making $\epsilon$ small enough can ensure that $x_{j}^* +
\epsilon (m_{j} - n_{j}) \geq 0$.

Before we proceed to checking the edge constraints, we make two
observations. Note that for any node $j$ in the tree, $j\in S\cap I$
then
\begin{itemize}
\item $x^*_j<1$, i.e. the mass $x^*_j$ put on $j$ by the \LP~ optimum
  $x^*$ is {\em strictly} less than 1. This is because of the
  alternating way in which the tree is constructed: a node $j$ in the
  tree is included in $S\cap I$ {\em only if} the parent $p$ of $j$ is
  in $S\cap I^*$ (note that the root $i\in S\cap I^*$ by
  assumption). However, from the definition of $I^*$, this means that
  $x^*_p>0$, i.e. the parent has positive mass at the \LP~ optimum
  $x^*$. This means that $x_j^* < 1$, as having $x_j^* = 1$ would mean
  that the edge constraint $x^*_p + x^*_j \leq 1$ is violated.
\item $j$ is not a leaf of the tree. This is because $S$ alternates
  between $I$ and $I^*$, and starts with $I^*$ at the root in level 1
  (which is odd). Hence $S\cap I$ will occupy even levels of the tree,
  but the tree has odd depth (by assumption $t$ is odd).
\end{itemize}
Now consider the edge constraints.  For any edge $(j,k)$, if the \LP~
optimum $x^*$ is such that the constraint is loose -- i.e. if $x^*_j +
x^*_k <1$ -- then making $\epsilon$ small enough will ensure that $x_j
+ x_k \leq 1$. So we only need to check the edge constraints which are
tight at $x^*$.

For edges with $x^*_j + x^*_k =1$, every time any copy of one of the
nodes $j$ or $k$ is included in $S\cap I$, the other node is included
in $S\cap I^*$. This is because of the following: if $j$ is included
in $S\cap I$, and $k$ is its parent, we are done since this means
$k\in S\cap I^*$. So suppose $k$ is not the parent of $j$. From the
above it follows that $j$ is not a leaf of the tree, and hence $k$
will be one of its children. Also, from above, the mass on $j$
satisfies $x^*_j <1$. However, by assumption $x^*_j + x^*_k =1$, and
hence the mass on $k$ is $x^*_k >0$. This means that the child $k$ has
to be included in $S\cap I^*$.

It is now easy to see that the edge constraints are satisfied: for
every edge constraint which is tight at $x^*$, every time the mass on
one of the endpoints is increased by $\epsilon$ (because of that node
appearing in $S\cap I$), the mass on the other endpoint is decreased
by $\epsilon$ (because it appears $S\cap I^*$). \hfill $\blacksquare$

\section{Max-product for Random Weighted Graphs}
\label{sec:two}

In this section, we establish the correctness and convergence of max-product 
algorithm when underlying graph is random and node weights are chosen as per 
exponential distribution. Specifically,  we consider two types of sparse random 
graphs, $G(n,c/n)$ and $G(n,r)$: 
\begin{itemize}
\item[{\bf 1.}] The $G(n,c/n)$ has $n$ nodes. An edge is present between any node-pair $i, j$ 
with  probability $c/n$ independently. Thus, on average $c(n-1)/2$ edges are present. 
\item[{\bf 2.}] The $G(n,r)$ has $n$ nodes. It is formed by sampling one of the $r$-regular $n$ node
graph uniformly at random. 
\end{itemize}
In either of these two cases, we assign node weight randomly. Specifically, let
$W_i$ denote the (random) weight of node $i$. Then, $W_i$ are independent and
identically distributed with exponential distribution of mean $1$. That is, for any
$\zeta \geq 0$, 
$$ \Pr\left( W_i \geq \zeta\right) = \exp\left(-\zeta\right). $$

\subsection{Results}

\noindent{\bf Convergence of max-product.} 
First, we establish that for $1-o(1)$ fraction of nodes, the algorithm
converges after {\em finitely} many iterations. Formally, we state
the result as follows.

\begin{theorem}\label{thm:main1}
Consider graph $G(n,c/n)$ or $G(n,r)$ with node weights assigned 
independently according  to exponential distribution of rate $1$. Let
$c \leq 2e$ and $r \leq 4$. Then, for  any $\epsilon > 0$, there exists
large enough $N(\epsilon)$ and $T(\epsilon)$ such that if 
$n \geq N(\epsilon)$, then following holds: for any node in $G(n,c/n)$ 
or $G(n,r)$,  say $i$, the $x_i^t (= x_i(\gamma^t))$ converges to the
correct value, $x_i^*$ with probability\footnote{Here, the probability 
distribution is induced by the choice of random graph and weights.}  
at least $1-\epsilon$ for $t \geq T(\epsilon)$. 
\end{theorem}

\noindent{\bf Correctness of max-product.} Theorem \ref{thm:main1} 
implies that almost all nodes converge to the correct solution. However,
questions that remain: (a) how to identify these `converged' nodes? 
and (b) do they have $1-o(1)$ fraction of weight of the MWIS? 
Indeed, we answer both of these questions in {\em affirmative}. 

We state the following simple stopping condition under which we 
will establish that all converged nodes get assigned the correct 
values, while all other nodes get assigned values $0$. As we shall 
establish, thus resulting assignment is indeed an independent
set with $1-o(1)$ fraction of weight of MWIS with high probability. 
Now the stopping condition and its approximation property.

\vspace{.1in}\noindent{\em Stopping condition.} At the end of iteration $t$, generate an estimate 
$\hat{I}^t$ using $\bx^t$ as $\{ i \in V: x_i^t = x_i^{t-1} =1 \}$.

\begin{theorem}\label{cor:main1}
Under the setup of Theorem \ref{thm:main1}, let algorithm stops after 
$t \geq T(\epsilon)$ steps producing $\bx^t$. Let $\hat{I}^t$ be 
the independent set obtained from $\bx^t$ as per the stopping condition 
described above. Then, $n \geq N_1(\beps)$ with large enough $N_1(\beps)$,
the weight of $\hat{I}^t$, $W(\hI^t)$ is such that
$$ \Pr\left(\frac{|W(\hI^t) - W(I^*)|}{W(I^*)} \geq \delta(\epsilon) \right) \leq \epsilon,$$
where $\delta(\epsilon) = O(\beps \log (1/\beps)) \to 0$ as $\epsilon \to 0$. 
Here, $I^*$ is the maximum weight independent set and $W(I^*)$ is its weight. 
\end{theorem}

\subsubsection{Outline of proof}

We now present a brief outline of the proof, and then present the
details in appendix due to space constraints.  Our results use the 
method of local weak convergence \cite{aldous}, and specifically 
the results of Gamarnik, Nowicki and Swirszcz \cite{GAM05}.  Under 
the random graph models considered here, i.e. random regular or 
Erdos-Renyi graph, for almost all nodes  of the graph their local 
neighborhood looks like a tree (see Lemma \ref{lem:shortcycle}). 
Now, under random selection of weights the assignment of node 
values under MWIS is determined by the local neighborhood for 
almost all nodes of the graph (see Lemma \ref{lem:four}) -- this property is
also known as the {\em correlation decay} property. Now, max-product
produces the correct estimate for each node with respect to its
computation tree. The computation tree of a node is equal to the local 
neighborhood as long as the local neighborhood is tree (recall Lemma \ref{lem:comp_tree}).  
Therefore, it is likely that for almost all nodes the max-product will produce
the correct estimate after finitely many iterations. In what follows,
we will make this precise by overcoming important technical subtlities. 
The correctness of stopping condition would follow from certain
``anti-monotonicity'' property of the max-product estimate procedure.
The good approximation property of the resulting estimate would follow
from certain extremality properties of Exponential distribution 
(see Lemma \ref{lem:extremal}).

\newpage

\appendix

\section{Proofs of Theorems \ref{thm:main1} and \ref{cor:main1}}


In this appendix, we present proofs of the two remaining Theorems 
\ref{thm:main1} and \ref{cor:main1}. We start with some useful
known properties. Then, we study certain ``local convergence''
properties of max-product on random graphs models considered
in this paper. Building upon these, we conclude the proof of
Theorem \ref{thm:main1}. Finally, using Theorem \ref{thm:main1}
and an extremal property of Exponential random variables 
we conclude the proof of Theorem \ref{cor:main1}. We note
that in this proof we assume that algorithm starts at time $t=0$
and not $t=1$ for a peculiar notational reason. Clearly, it is
of no non-trivial relevance.

\subsection{Useful properties}

Here, we describe useful definitions, notations and properties for proving 
Theorem \ref{thm:main1}. To this end, consider a fixed node $i$ in graph $G$.
Define, $V_i(t) \subset V$ as 
$$ V_i(t) = \{ j \in V : \text{there is a path between $i$ and $j$ of length no more than $t$}\}.$$
Let $E_i(t) \subset E$ be set of edges incident between these vertices and $G_i(t) = (V_i(t), E_i(t))$
be the subgraph of $G$ thus created. As defined earlier, let $T_i(t)$ be the computation
tree of node $i$ till iteration $t$. Then, it is straightforward that $T_i(t) = G_i(t)$ (in terms of 
graph structure only) if $G_i(t)$ is itself a tree.  

Some notation and definitions before stating result about structural properties of 
$G_i(t)$ when $G = G(n, c/n)$ or $G=G(n,r)$. A Poisson tree of depth $t$ and parameter $c$, denoted as $\cT(c,t)$, is
constructed as follows: starting with root, say $0$, add Poisson($c$) number of children to it. 
Recursively,  for each of thus created children, add  Poisson($c$) number of children independently
till the tree has depth $t$. A regular tree of depth $t$ and parameter $r$, denoted as $\bT(r,t)$,
is constructed as follows: starting with root, say $0$, add $r$ number of children to it. Recursively,
for each of thus created children, add $r-1$ number of children till the tree has depth $t$.  

Now, we state the following well-known (and very important for us) property about the 
local structure of $G(n,c/n)$ and $G(n,r)$  (see \cite{Cy1} and \cite{Cy2} respectively for details).

\begin{lemma}\label{lem:shortcycle}
Consider graph $G = G(n, c/n)$ or $G=G(n, r)$ with finite values of $c$ and $r$. Consider
a fixed node $i$ (numbering of nodes is done prior to selection of edges). Then, as $n\to \infty$,
\begin{itemize}
\item[(a)] For $G=G(n,c/n)$, the $G_i(t)$ converges (in distribution) to the Poisson tree, $\cT(c,t)$;
\item[(b)] For $G = G(n,r)$, the $G_i(t)$ converges (in distribution) to the regular tree, $\bT(r,t)$.
\end{itemize}
\end{lemma}

\subsection{Local convergence of max-product}

Consider the max-product algorithm running at a particular node, say $i$. In what follows,
$i$ is always used to denote a fixed node\footnote{This fixed node $i$ is chosen a priori selection of random graph
structure and weights or equivalently, its selection is done uniformly at random from $n$ nodes for the purposes
of the proofs.}. We wish to understand the evolution its estimate $x_i^t$, which 
depends on the messages $\gamma^t$. Specifically, as we stated earlier, the 
messages $\gamma^t$ can be defined recursively on the computation tree $T_i(t)$
as follows: node $k$ generates message for its parent, say $p(k)$, using
messages from its children set, say $c(k)$ and its weight $W_k$ as
$$ \gamma_{k\to p(k)}^t = \max\left(0, W_k - \sum_{\ell \in c(k)} \gamma^t_{\ell\to k}\right).$$
When, $c(k) = \emptyset$ that is $k$ is a leaf node of $T_i(t)$, then due to $0$ initial condition, 
$$ \gamma_{k\to p(k)}^t = \max\left(0, W_k \right) ~=~W_k,  $$
since $W_k \geq 0$ with probability $1$.  Call the $\gamma^t$ obtained with this $0$
initial conditions as $\gamma^t(\bzero)$.  

Now, suppose initial condition at a leaf node $k$, that is at depth $t$ of the 
computation tree, be $W_k$ instead of $0$. Equivalently, if incoming messages to
$k$ (at the depth $t$) summed upto $\geq W_k$, then   
$$ \gamma_{k\to p(k)}^t = \max\left(0, W_k -W_k \right) ~=~0.  $$
Call the $\gamma^t$ obtained along all the edges of computation tree $T_i(t)$
with such initial condition as $\gamma^t(\bW)$: that is, all leaf nodes at 
level $t$ of $T_i(t)$ have initial condition equal to their node weights, while
leaf nodes\footnote{Such leaf nodes can exists for Poisson
tree, but will not exist for a regular tree.} at level $< t$ have the usual initial condition $0$. 
  
Finally, consider starting algorithm with initial condition $L_k\in [0, W_k]$ at leaf
node $k$ at depth $t$ in $T_i(t)$. Then, the messages from leaf nodes, 
say $k$ of $T_i(t)$, is 
$$0 \leq  \gamma_{k\to p(k)}^t = \max\left(0, W_k - L_k \right) \leq W_k.$$
Let us denote the message obtained by starting with initial condition $\bL$ (vector
of initial condition values for leaf nodes at depth $t$ of $T_i(t)$) as
$\gamma^t(\bL)$.  

The above discussion implies the following: for $t = 1$, for any $j$ which is
children of $i$ and any starting condition $\bzero \leq \bL \leq \bW$ (component-wise
inequality), 
$$ \gamma^t_{j\to i}(\bzero) \geq \gamma^t_{j\to i}(\bL) \geq \gamma^t_{j\to i}(\bW).$$
This non-increasing behavior of messages received at root node as initial condition 
is increasing holds true for all odd $t$ inductively (can be easily verified).  
\begin{lemma}\label{lem:monotone}
Consider an odd $t$, fixed node $i$ and its computation tree $T_i(t)$. 
Then, for any non-negative starting condition (for leaf nodes at depth $t$) 
$\bzero \leq \bL \leq \bW$ (component-wise) and any children $j$ of root node $i$, 
$$ \gamma^t_{j\to i}(\bzero)\geq \gamma^t_{j\to i}(\bL) \geq \gamma^t_{j\to i}(\bW). $$
\end{lemma}
Next, we consider the estimation of node $i$ based on its messages at time $t$.
For convenience, we define notion of bonus at node $i$ at time $t$, denote
as $B_i^t$ as follows: 
$$ B_i^t = W_i - \sum_{j\in \cN(i)} \gamma^t_{j\to i}.$$
As per max-product algorithm, the estimation $x_i^t = 1$ if 
$B_i^t > 0$; $x^t_i = 0$ if $B_i^t < 0$ and $?$ otherwise. 
Let $B_i^t(\bzero), B_i^t(\bL)$ and $B_i^t(\bW)$ denote 
bonus values when algorithm is started with initial conditions
$\bzero \leq \bL \leq \bW$ for leaf nodes of
$T_i(t)$ respectively.  From Lemma \ref{lem:monotone} and definition
of bonus, it follows that for any odd $t$,
\begin{eqnarray}
B_i^t(\bzero)  & \leq &  B_i^t(\bL) ~\leq~ B_i^t(\bW). \label{eq:bonus} 
\end{eqnarray}
Due to very similar reasoning, it also follows that the bonuses have
anti-monotone property starting with $\bzero$ initial condition. That is,
for any odd $t$
\begin{eqnarray}
B_i^t(\bzero)  & \leq &  B_i^{t+1}(\bzero). \label{eq:bonus1} 
\end{eqnarray}
The following result is direct adaption of  \cite[Theorem 8]{GAM05}. It will be 
essential to complete our result. 
\begin{lemma}\label{lem:four}
Consider a fixed $\phi \in [0,\infty)$ and initial conditions $\bzero \leq \bL \leq \bW$
for leaf nodes at depth $t$ of $T_i(t)$.  Let $\beps > 0$ be given. Then, there exists an odd $t(\beps)$ large enough
such that the following holds: for $t = t(\beps)$, 
\begin{itemize}
\item[(a)] Let, $G_i(t)$ $($and hence $T_i(t)$$)$ be (distributed as) Poisson tree $\cT(c,t)$ with $c \leq 2e$. Then, 
$$ \Pr\left(B_i^t(\bzero) \leq \phi\right) \leq \Pr\left(B_i^t(\bL) \leq \phi\right) 
\leq \Pr\left(B_i^t(\bW) \leq \phi\right) \leq \Pr\left(B_i^t(\bzero) \leq \phi\right) + \beps. $$

\item[(b)] Let, $G_i(t)$ $($and hence $T_i(t)$$)$ be (distributed as) regular tree $\bT(r,t)$ with $r \leq 4$. Then, 
$$ \Pr\left(B_i^t(\bzero) \leq \phi\right) \leq \Pr\left(B_i^t(\bL) \leq \phi\right) \leq \Pr\left(B_i^t(\bW) \leq \phi\right) \leq \Pr\left(B_i^t(\bzero) \leq \phi\right) + \beps. $$

\end{itemize}
\end{lemma}

\subsection{Proof of Theorem \ref{thm:main1}: putting things together}

Now we complete the proof of Theorem \ref{thm:main1} using Lemmas \ref{lem:shortcycle}, 
\ref{lem:monotone}  and  \ref{lem:four} along with some properties of the algorithm. We
state proof for $G(n,c/n)$ with $c\leq 2e$ and $G(n,r)$ with $r\leq 4$ simultaneously. 

Consider a fixed node $i$ of $G$, where $G=G(n,c/n), c \leq 2e,$ or $G=G(n,r), r \leq 4$.
Let $\beps > 0$ be given. Let $E_1$ be the event that the local neighborhood of depth 
$t$ of node $i$, $G_i(t)$ is {\em tree}. We will assume some odd $t = t(\beps)$ or $t(\beps)+1$
as required per Lemma \ref{lem:four}. As part of the algorithm, the bonus values at 
node $i$, $B_i^t(\bzero)$ can be computed recursively starting with $\bzero$ initial
condition at the leaf nodes (at depth $t$) of its computation tree $T_i(t)$
as described before. Now, consider the bonus value $B_i^{t+1}(\bzero)$ at the
{\em next} step. Due to the update equation of our algorithm and 
computation tree structure, $T_i(t) \subset T_i(t+1)$, it
is easy to see that  $B_i^{t+1}(\bzero)$ is equal to $B_i^t(\bL)$ for some 
initial condition $\bL$, $\bzero \leq \bL \leq \bW$, for leaf nodes at depth $t$
of $T_i(t)$. This is because irrespective of incoming messages, the 
message generated along any edge is always between $[0, W]$, where $W$ is
the node weight. Similarly, we can inductively argue that $B^{s}_i(\bzero)$ is 
equal to $B_i^t(\bL_s)$ for some $0\leq \bL_s \leq  \bW$ for all $s \geq t$. Therefore, 
using Lemma \ref{lem:monotone} and its consequence (\ref{eq:bonus}),  
for any $s \geq t$, 
\begin{eqnarray} 
B_i^t(\bzero) & \leq &  B_i^s(\bzero)  ~\leq~ B_i^t(\bW). \label{eqp:1}
\end{eqnarray}
Suppose, that $E_1$ is true. Then, $G_i(t) = T_i(t)$ and the $W_i$ 
is independent of messages $\gamma^t$ based on computation tree 
$T_i(t)$. Therefore, $B_i^t(\bzero) \neq 0$ and $B_i^t(\bW) \neq 0$ 
with probability $1$ due to $W_i$ being drawn from a continuous distribution. 
Therefore, by definition of $x_i^s$ based on $B_i^s(\bzero)$ and 
(\ref{eqp:1}), it follows that  under event $E_2 = \{B_i^t(\bzero) > 0 \}$, 
the $x_i^s = 1$ for all $s \geq t$. Consider event $E_3 = \{B_i^t(\bW) < 0 \}$.
Given $E_1$, $E_3 = \{B_i^t(\bW) \leq 0 \}$ since $B_i^t(\bW) \neq 0$ with 
probability $1$. Therefore, using similar reasoning we have that $x_i^s = 0$
for all $s \geq t$. From this discussion, it follows that for $s \geq t$
\begin{eqnarray}
\Pr\left(x_i^s  ~\mbox{converges}\right) & \geq & \Pr((E_2 \cup E_3) \cap E_1) ~=~{\Pr}_{G_i(t)}((E_2 \cup E_3) \cap E_1), \label{eqp:x0}
\end{eqnarray}
where in the last equality we introduce the sub-script $G_i(t)$ as the probability of event $(E_2 \cup E_3) \cap E_1$ 
primarily depends only on local neighborhood of depth $t$, $G_i(t)$. Here, 
by $\Pr_{G_i(t)}$ we mean the distribution induced by the random graph $G(n,c/n)$ or
$G(n,r)$ and the random node weights on the local neighborhood of node $i$
upto depth $t$. Inspired by Lemma \ref{lem:shortcycle},  consider the distribution 
induced by Poisson tree $\cT(c,t)$ or regular tree $\bT(r,t)$ (depending upon
random graph model of $G$) and the independent random node weights -- 
denote this by $\bQ(\cdot)$. 

Now, the sequence (dependent on number of nodes $n$ in $G$) of 
distributions $\Pr_{G_i(t)}(\cdot)$ converges to $\bQ(\cdot)$ as
per Lemma \ref{lem:shortcycle}. This convergence is defined over
appropriate topology of local weak convergence (see, \cite{Aldous01}
for example). Now, it can be checked that 
$(E_2 \cup E_3)\cap E_1$ is an open set. Therefore, by Portmanteau 
theorem it follows that 
\begin{eqnarray}
 \lim\inf_{n\to\infty} {\Pr}_{G_i(t)}((E_2 \cup E_3) \cap E_1) & \geq & \bQ((E_2 \cup E_3) \cap E_1). \label{eqp:x1}
\end{eqnarray}
Equivalently, for selection of $N(\beps)$ large enough we have that when $n \geq N(\beps)$
\begin{eqnarray}
 {\Pr}_{G_i(t)}((E_2 \cup E_3) \cap E_1) & \geq & \bQ((E_2 \cup E_3) \cap E_1) - \beps. \label{eqp:x11}
\end{eqnarray}
Therefore, in order to establish that $x_i^s$ converges for $s \geq t$ with  enough
probability it is sufficient to show that 
$$ \bQ((E_2 \cup E_3) \cap E_1) \geq 1-\beps.$$
Now consider the following. 
\begin{eqnarray}
\bQ((E_2 \cup E_3) \cap E_1) & \stackrel{(a)}{=} & \bQ(E_2 \cup E_3) \nonumber \\
& = & \bQ(E_2) +  \bQ(E_3 \cap E_2^c) \nonumber \\
   & \stackrel{(b)}{=} & \bQ(E_2) + \bQ(E_3) \nonumber \\
   & \stackrel{(c)}{=} & \bQ\left(B_i^t(\bzero) > 0 \right) +  \bQ\left(B_i^t(\bW) \leq 0 \right) \nonumber \\
   & = & 1 - \bQ\left(B_i^t(\bzero) \leq 0 \right) +  \bQ\left(B_i^t(\bW) \leq 0 \right)   \nonumber \\
   & \geq & 1 - \epsilon, \label{ef3} 
 \end{eqnarray}
where (a) follows from fact that under $\bQ$ the $E_1$ is always true; (b) follows 
from --
$E_3 = \{ B^t_i(\bW) < 0 \} \subset \{ B^t_i(\bzero) \leq 0 \} = E_2^c$; (c) follows
from definition and (\ref{ef3}) follows from Lemma \ref{lem:four} (for either $\cT(c,t)$ 
or $\bT(r,t)$). Thus, (\ref{eqp:x0}), (\ref{eqp:x1}) and (\ref{ef3}) imply that 
$x_i^s$ converges with probability at least $1-2\beps$ for $n \geq N(\beps)$ and
$s \geq t = t(\beps)$. 

To prove the correctness of $x^t_i$ upon convergence, note the following: 
by Theorem \ref{thm:lp_mp} and it follows that if $x_i^t$ has converged to $0$
or $1$ then the corresponding LP optimum solution must have those assignment
for node $i$. By Lemma \ref{lem:partial} it follows that these are the 
assignment from the MWIS. Therefore, if $x_i(\gamma^s)$ converges (or in
fact is equal to 0, or 1, for two subsequent time steps) then Theorem
\ref{thm:lp_mp} states that $x_i(\gamma^s) = x^*_i$, the value of the
unique \LP~ optimum for $i$ and Lemma \ref{lem:partial} then implies that
this estimate is actually correct for $G$. Therefore, we have proved that 
the converged values is the correct value. This completes the proof of 
Theorem \ref{thm:main1} by re-selection $\beps = \beps/2$ and selection
appropriate values for $N(\beps)$ and $T(\beps)$.

\subsection{Proof of Theorem \ref{cor:main1}}

We need to establish that (a) the stopping condition induces an
independent set, $\hI^t$ and (b) the $\hI^t$ has high enough
weight. To this end, recall that if $x_i^{t-1} = x_i^t = 1$ 
then by Theorem \ref{thm:lp_mp} it must be that the LP optimum
assignment has $x_i^* = 1$ and by Lemma \ref{lem:partial} it is
indeed equal to the assignment as per the MWIS. Since $\hI^t$ 
contains only such nodes, it follows that $\hI^t \subset I^*$
and hence a valid independent set. 

Now, by Theorem \ref{thm:main1} it follows that for graph with 
$n \geq N(\beps)$ and number of iterations $t \geq T(\beps)$, at
least $1-\beps$ fraction of nodes find their right assignment with
probability at least $1-\beps$. Therefore, it follows that 
$| I^* \backslash \hI^t | \leq \beps n$ with probability at least
$1-\beps$. Now, consider the following {\em extremal} property of
Exponential random variables. 
\begin{lemma}\label{lem:extremal}
Consider $n$ i.i.d. random variables $X_1,\dots,X_n$ with Exponential
distribution of mean $1$. Let the ordered sequence be 
$X_{\pi(1)} \geq \dots \geq X_{\pi(n)}$. Then, for any $\beps \in (0,e^{-10})$
$$ \Pr\left( \frac{1}{n}\sum_{i=1}^{\beps n} X_{\pi(i)}  \geq 
2\beps (1+\ln (1/\beps))  \right) \leq  \exp\left(- n \beps \ln (1/\beps)\right).
$$
\end{lemma}
\begin{proof}
The proof follows by an application of Cramer's Theorem for Exponential
random variables. Specifically, given $N$ i.i.d. random variables $Y_1,\dots, Y_N$
with mean $1$ and Exponential distribution, for any $L \geq 10$ it
follows that 
\begin{eqnarray}
\Pr\left( \frac{1}{N} \sum_{k=1}^N Y_k \geq L \right) & \leq & \exp\left(- N(L-1 - \log L) \right) \nonumber \\
& \leq & \exp\left(-NL/2\right).
\label{eq:cn1}
\end{eqnarray}
Therefore, by application of (\ref{eq:cn1}) it follows that for any collection of $\beps n$
of $X_1,\dots, X_n$, their summation is no larger than $n \beps L$ with probability at least
$1 - \exp(-n\beps L/2)$ for $L \geq 10$. 

Now, given $n$ random variables there are ${n \choose \beps n}$ distinct ways to
select $\beps n$ indices. Therefore, by an application of union bound, above 
discussion and Striling's approximation it follows that 
\begin{eqnarray}
\Pr\left( \frac{1}{n}\sum_{i=1}^{\beps n} X_{\pi(i)}  \geq \beps L \right) & \leq & 
{n \choose \beps n} \exp\left(-\frac{n\beps L}{2}\right) \nonumber \\
& \leq &  \exp\left(n \beps \ln (1/\beps) + n\beps - n\beps L/2\right). \label{eq:cn2}
\end{eqnarray}
Select $L = L(\beps) = \max\{2 (1+ \ln (1/\beps)), 10\}$. Then, we obtain that 
\begin{eqnarray}
\Pr\left( \frac{1}{n}\sum_{i=1}^{\beps n} X_{\pi(i)}  \geq \beps L(\beps)  \right) & \leq & 
\exp\left(-n \beps \ln (1/\beps)\right). \label{eq:cn3}
\end{eqnarray}
Note that for $\beps \leq e^{-10}$, $L(\beps) =  2 (1+ \ln (1/\beps)) \geq 10$. 
This completes the proof of Lemma \ref{lem:extremal}. \eproof 
\end{proof}

From Lemma \ref{lem:extremal}, it follows that for $n$ large enough (say
larger than $N_1(\beps)$) the 
the net weight of nodes in $|I^* \backslash \hI^t|$ is at most  $\gamma(\beps) n$
with probability at least $1-\beps$ for $\gamma(\beps) = O(\beps \ln (1/\beps))$ as 
$\beps \to 0$. Clearly, $\gamma(\beps) \to 0$ as $\beps \to 0$.  As established 
in \cite{GAM05}, the weight of the maximum weight independent set is 
$\Theta(n)$ with probability $1-o(1)$ under the setup of our interest. 
It follows (using union bound) that for $N_1(\beps)$ large enough, for $n \geq N_1(\beps)$,
\begin{eqnarray}
P\left(  \frac{|W(\hI^t) - W(\cI^*)|}{W(\cI^*)}  \geq  \delta(\epsilon) \right)  & \leq  & 2\epsilon,
\end{eqnarray}
with $\delta(\beps) = O(\beps \log (1/\beps))$. This completes the proof of Theorem \ref{cor:main1}.

\end{document}